\documentclass{article}
\def\a{\alpha}
\def\b{\beta}
\def\g{\gamma}

\def\h{\eta}

\def\la{\lambda}
\def\m{\mu}

\def\vp{\varphi}

\def\ps{\psi}

\def\P{\Phi}
\def\Ps{\Psi}
\def\La{\Lambda}
\newcommand{\C}{\mathbb C}

\newcommand{\Ical}{{\cal I}}

\newcommand{\Qcal}{{\cal Q}}
\newcommand{\Xcal}{{\cal X}}
\newcommand{\Ucal}{{\cal U}}
\newcommand{\Acal}{{\cal A}}

\def\>{\rangle}
\def\<{\langle}
\def\={\ =\ }
\def\+{\dagger}
\def\e{\textrm{e}}
\def\ii{\textrm{i}}
\def\N2{$N\,{=}\,2$}
\def\Ng4{$N\,{=}\,4$}
\def\pa{\partial}
\def\diff{\mbox{d}}
\def\dt{\widetilde{\mbox{d}}}

\def\sfrac#1#2{{\textstyle\frac{#1}{#2}}}
\newcommand{\bsy}[1]{\boldsymbol{#1}}
\newcommand{\ov}[1]{\overline{#1}}
\newcommand{\wt}[1]{\widetilde{#1}}
\newcommand{\lb}{\bar{\la}}
\newcommand{\mb}{\bar{\mu}}
\newcommand{\Psb}{\ov{\Psi}}
\newcommand{\Tb}{\ov{T}}

\newcommand{\At}{\wt{A}}
\newcommand{\Pst}{\wt{\Psi}}
\def\beq{\begin{equation}}
\def\eeq{\end{equation}}
\def\bea{\begin{eqnarray}}
\def\eea{\end{eqnarray}}
\def\1ad{\mbox{\normalsize $^1$}}
\def\2ad{\mbox{\normalsize $^2$}}
\def\3ad{\mbox{\normalsize $^3$}}
\def\4ad{\mbox{\normalsize $^4$}}
\def\5ad{\mbox{\normalsize $^5$}}
\def\6ad{\mbox{\normalsize $^6$}}
\def\7ad{\mbox{\normalsize $^7$}}
\def\8ad{\mbox{\normalsize $^8$}}
\def\makefront{
\vspace*{1cm}\begin{center}
\def\sp{
\renewcommand{\thefootnote}{\fnsymbol{footnote}}
\footnote[4]{corresponding author : \email_speaker}
\renewcommand{\thefootnote}{\arabic{footnote}}
}
\def\newtitleline{\\ \vskip 5pt}
{\Large\bf\titleline}\\
\vskip 1truecm
{\large\bf\authors}\\
\vskip 5truemm
\addresses
\end{center}
\vskip 1truecm
{\bf Abstract:}
\abstracttext
\vskip 1truecm
}
\usepackage{amssymb}
\usepackage{amsmath}
\usepackage{amscd}
\usepackage{latexsym}
\usepackage{ifthen}
\begin {document}                 

\begin{flushright}      
hep-th/0212335\\        
ITP--UH--32/02\\        
\end{flushright}        

\def\titleline{
Solving String Field Equations:
\newtitleline
New Uses for Old Tools
}
\def\email_speaker{
{\tt 
lechtenf@itp.uni-hannover.de
}}
\def\authors{
A. Kling,
O. Lechtenfeld,
A.D. Popov, 
S. Uhlmann
}
\def\addresses{
Institut f\"ur Theoretische Physik, Universit\"at Hannover\\
Appelstra\ss{}e 2, D--30167 Hannover, Germany\\
Email: kling, lechtenf, popov, uhlmann@itp.uni-hannover.de
}
\def\abstracttext{
This is the contents of a talk by O.~L.~presented   
at the 35th International Symposium Ahrenshoop      
in Berlin, Germany, 26--30 August 2002.             
It is argued that the (NS-sector) superstring field equations are integrable,
i.~e.~their solutions are obtainable from linear equations.
We adapt the 25-year-old solution-generating ``dressing'' method
and reduce the construction of nonperturbative superstring configurations
to a specific cohomology problem. The application to vacuum superstring
field theory is outlined.
}
\large
\makefront


\section{Zero-curvature and linear equations (old tools)}

The flatness of a gauge connection~$A$,\footnote{
We suppress the spacetime coordinate dependence throughout.}
\beq \label{flat0}
F(A)\ \equiv\ \diff A+A^2\=(\diff+A)^2\=0 \quad,
\eeq
may be seen as the compatibility condition for a linear system:
\beq \label{linsys0}
\exists\Ps\quad\textrm{with}\quad (\diff+A)\,\Ps\=0
\qquad\Longrightarrow\qquad F(A)=0 \quad.
\eeq
If we take the auxiliary function~$\Ps$ to be Lie-group valued,
solutions of the linear system yield flat connections $A=\Ps\diff\Ps^{-1}$
which are, however, pure gauge and hence trivial.

The situation changes when $\diff$ is just a {\it partial\/} differential 
and one has a {\it second\/} partial differential~$\dt$ 
which anticommutes with the former.
We may then combine the two and also their corresponding partial connections
$A$ and~$\At$ to a {\it family\/} 
\beq
A(\la)\=\At+\la A \qquad\textrm{and}\qquad \diff(\la)\=\dt+\la\diff 
\qquad\textrm{with}\quad \dt^2={\diff\big.}^2=\dt\diff+\diff\dt=0 \quad,
\eeq
where some parameter $\la{\in}\C P^1$ is introduced.
The {\it extended\/} zero-curvature equation reads
\beq
0 \= F\bigl(A(\la)\bigr) \= \bigl( \diff(\la)+A(\la) \bigr)^2 \= 
  ( \dt\At {+} \At^2 )
+ \la ( \dt A {+} \diff\At {+} \{ A,\At \} )
+ \la^2 ( \diff A {+} A^2 ) \quad.
\eeq
Exploiting the gauge freedom herein allows one to gauge away\footnote{
on a topologically trivial manifold} 
one of the two partial connections. We fix $\At=0$ and remain with
\beq \label{flat}
\dt A \= 0 \qquad\textrm{and}\qquad \diff A+A^2 \= 0 \quad.
\eeq
As long as the $\diff$- and $\dt$-cohomologies are empty, either one of
these two equations is solved by the introduction of a prepotential, 
for which the remaining equation imposes a second-order relation: 
\bea
& A\=\dt\Upsilon \qquad\Longrightarrow\qquad
  \diff\dt\Upsilon + (\dt\Upsilon)^2 \=0 & \quad, \\
& A\=\e^{-\P}\diff\e^{\P} \qquad\Longrightarrow\qquad
  \dt ( \e^{-\P}\diff\e^{\P} ) \=0 & \quad.
\eea
The first and second of these reductions go back to 
\cite{Leznov:mx} and \cite{Yang:1977zf}, respectively.
Despite appearance, $A$ is {\it not\/} pure gauge unless $\dt\e^{\P}=0$
in which case $\e^{\P}$ qualifies as a gauge parameter compatible 
with $\At=0$.

The extended linear system associated with (\ref{flat}) is
\beq \label{linsys}
\bigl( \dt + \la\diff + \la A \bigr)\,\Ps(\la)\=0 \quad.
\eeq
Due to the $\la$-dependence, 
it gives rise to nontrivial solutions of~(\ref{flat}).
Indeed, if $\Ps$ does not depend on~$\la$ 
the partial connection~$A$ must be pure gauge:
\beq
\Ps(\la)\= \e^{-\La}
\qquad\Longrightarrow\qquad
\dt\,\e^{-\La}\=0\=(\diff+A)\,\e^{-\La} \quad.
\eeq
The compactness of $\C P^1$ excludes nontrivial holomorphic~$\Ps(\la)$;
hence we consider {\it meromorphic\/}~$\Ps(\la)$.
Not allowing for poles at $\la{=}0$ or $\la{=}\infty$, we fix the asymptotics
\beq
\Ps(\la)\quad\longrightarrow\quad\begin{cases}
1 - \la\,\Upsilon + O(\la^2)  & \textrm{for} \quad \la\to0 \\
\e^{-\P} + O(\sfrac{1}{\la}) & \textrm{for} \quad \la\to\infty \end{cases}
\eeq
by a convenient residual gauge choice. 
The form of~$\Ps(\la)$ is constrained further by the antihermiticity of~$A$
up to a gauge transformation.
Hermitian conjugation extends to an involution $\Ps\mapsto\Psb$ which sends 
$\diff\mapsto-\dt$ and $\dt\mapsto\diff$ but $\la\mapsto\lb$.
With this, an antihermitian connection requires that
\beq \label{reality}
\e^{-\P}\= \Ps(\la)\,\ov{\Ps(-1/\lb)} \quad.
\eeq
Finally, we may reconstruct $A$ from a given solution~$\Ps(\la)$ via
\beq \label{solveA}
A\= \Ps(\la) \bigl( \diff + \sfrac{1}{\la}\dt \bigr) \Ps(\la)^{-1} \quad.
\eeq

\section{Single-pole ansatz}

The simplest non-constant meromorphic function possesses a single pole.
The corresponding ansatz,\footnote{
This is an essential building block in the ``dressing method'', 
a solution-generating technique invented by \cite{Zakharov:pp,zakh2} 
and developed by \cite{Forgacs:1983gr}.}
\beq \label{ansatz}
\Ps(\la)\=\bsy{1}-\frac{\la(1{+}\m\mb)}{\la-\m}\,P \quad,
\eeq
contains a moduli parameter~$\m$ (the location of the pole) and a
Lie-group valued $\la$-independent function~$P$ to be determined.

It turns out that all information resides in eqs.~(\ref{reality}) 
and~(\ref{solveA}). Since their left hand sides are independent of~$\la$,
the poles at $\la{=}\m$ and $\la{=}{-}1/\mb$ of their right hand sides
must be removable. Putting to zero the residues in (\ref{reality}) yields
algebraic relations,
\beq
P^2 \= P \= \ov{P}  \qquad\Longleftrightarrow
\qquad\textrm{$P$ is a hermitian projector} \quad.
\eeq
As such, $P$ can be parametrized with a ``column vector'' $T$ via
\beq
P\=T\,\sfrac{1}{\Tb T}\,\Tb \quad.
\eeq
Similarly, the vanishing residues in (\ref{solveA}) produce 
differential equations,
\bea
& &\quad P\,(\dt+\m\diff)P \= 0 \= (\bsy{1}{-}P)(\diff-\mb\dt)P \\
&\Longleftrightarrow &\quad (\bsy{1}{-}P)(\diff-\mb\dt)\,T\=0 \\
&\Longleftarrow &\quad (\diff-\mb\dt)\,T\= 0 \quad,
\eea
whose solution requires the analysis of the cohomology of the operator
$\diff{-}\mb\dt$.\footnote{
More generally, it suffices to solve the `eigenvalue' equation
$(\diff-\mb\dt)T=T\,\a$ with a flat connection $\a$, but we can
gauge $\a$ to zero locally.}
Hence, the original nonlinear problem~(\ref{flat}) has been reduced
(``linearized'') to a linear homogeneous equation for~$T$. Any nonsingular 
element in the kernel of~$\diff{-}\mb\dt$ yields a projector~$P$,
through which the prepotentials and the connection are expressed as follows,
\bea
&& \Upsilon\=-\sfrac{1{+}\m\mb}{\m}\,P \qquad\textrm{and}\qquad
   \e^{-\P}\=\bsy{1}-(1{+}\m\mb)\,P  \quad, \\
&& A\=-\sfrac{1{+}\m\mb}{\m}\,\dt P 
    \=(1{+}\m\mb)\,P\diff P
     -(1{+}\sfrac{1}{\m\mb})(1{-}P)\diff P \quad.
\eea

\section{String fields (new uses)}

The complete structure presented so far carries over almost verbatim 
to string field theory. In addition to being Lie-group or Lie-algebra valued,
our objects now are interpreted as string fields and are to be multiplied
via Witten's star product~\cite{Witten:1985cc} (which we suppress).
More precisely, the unextended situation~(\ref{flat0}) and~(\ref{linsys0})
corresponds to the cubic open bosonic string~\cite{Witten:1985cc}, 
with $\diff\mapsto Q$ (the BRST operator) and $A$ having ghost number one.
The linear system $(Q{+}A)\Ps=0$ yields only trivial solutions to the
string field equation of motion, $QA{+}A^2=0$.

Surprisingly, the extended case (\ref{flat}) and (\ref{linsys}) can be
mapped onto the NS sector of cubic open {\it superstring\/} field theory
\cite{Witten:1986qs}, by $\diff\mapsto Q$ and $\dt\mapsto\h_0$.
Here, $\h_0$ denotes the graded commutator action of the zero mode of $\h$,
which emerges from bosonizing the worldsheet supersymmetry ghosts via
$\g=\h\,\e^{\vp}$ and $\b=\e^{-\vp}\pa\xi$.
The NS string field~$A$ carries ghost number one and picture number zero
and lives in the ``large Hilbert space'' (including~$\xi_0$).
Consequentially, its equations of motion~\cite{Preitschopf:fc,Arefeva:1989cp}
are 
\beq \label{eomcubic}
\h_0\,A\=0 \qquad\textrm{and}\qquad Q A+A^2\=0 \quad.
\eeq
Berkovits' nonpolynomial open superstring (in the NS sector) 
\cite{Berkovits:1995ab} is also found:
\beq
A\=\e^{-\P}Q\e^{\P} \qquad\Longrightarrow\qquad
\h_0 ( \e^{-\P}Q\e^{\P} ) \=0 \quad.
\eeq
The Berkovits string field~$\P$ has vanishing ghost and picture numbers.

It is known that $\h_0$ and $Q$ have zero cohomology in the
large Hilbert space. Therefore, we can get all solutions to~(\ref{eomcubic})
from a ``linear superstring system''~\cite{Lechtenfeld:2002cu},
\beq \label{key}
\bigl( Q + \sfrac{1}{\la}\h_0 + A \bigr)\,\Ps(\la) \=0 \quad.
\eeq
This is the key equation for generating nonperturbative classical 
superstring field configurations.  Repeating the earlier analysis, 
the analog of the ansatz (\ref{ansatz}) produces a hermitian projector
string field~\footnote{
The identity string field $\Ical$ is the unit element in Witten's star algebra.}
\beq
P\=T\,\sfrac{1}{\Tb T}\,\Tb 
\qquad \textrm{subject to}\ \qquad
(\Ical{-}P)(Q-\mb\h_0)P \=0 \quad.
\eeq
A sufficient condition for the latter equation to hold is
\beq
(Q-\mb\h_0)\,T \= 0 \quad.
\eeq
{}From a solution~$T$ one builds $P$ and finally (using star multiplication)
one reconstructs~\cite{Lechtenfeld:2002cu} the string fields
\beq
\e^{-\P}\=\Ical-(1{+}\m\mb) P \qquad\textrm{and}\qquad
A\=-\sfrac{1{+}\m\mb}{\m}\,\h_0 P \quad.
\eeq

\section{Ghost-picture modification}

Our ``master equation''~(\ref{key}) has a flaw: 
It is inhomogeneous in picture number because $\h_0$ lowers the picture 
by one unit. Therefore, any solution~$\Ps$ is an {\it infinite sum\/}
over all picture sectors (requiring an extension of standard superstring
field theory), unless we modify our equation by introducing a picture-raising
multiplier,
\beq
\h_0 \quad\rightarrow\quad X(\ii)\,\h_0
\qquad\textrm{with}\qquad X(\ii)\=\{Q,\xi(\ii)\} \quad,
\eeq
to be inserted at the string midpoint (with worldsheet coordinate~$\ii$).
The master equation then changes to
\beq \label{modkey}
(Q + \sfrac{1}{\la} X(\ii)\,\h_0 + A)\,\Ps(\la)\=0 \quad,
\eeq
and the string field equations of motion become
\beq
QA + A^2 \= 0 \qquad\textrm{and}\qquad
X(\ii)\,\h_0\,A \= X(\ii)\,\h_0 ( \e^{-\P}Q\e^{\P} ) \=0 \quad.
\eeq
The modification may create extra solutions due to zero modes of~$X(\ii)$.

\section{Shifting the background}

We may view solving~(\ref{modkey}) as ``dressing'' the vacuum solution,
\beq
(\Ps_0,A_0) = (\Ical,0) \qquad\longmapsto\qquad (\Ps,A) \quad,
\eeq
via $\ \Ps=\Ps(\la)\Ps_0\ $ and $\ A={\rm Ad}_{\Ps(\la)}A_0\ $.
This process can be iterated, and such transformations form a group.
In fact, they generate all solutions and, hence, relate all classical
backgrounds with one another. Clearly, shifting the vacuum background
$(\Ps_0,A_0)$ to a new reference $(\Ps_1,A_1)$ is also 
a dressing transformation:
\beq 
\parbox{4cm}{
    \textrm{background:}
    \phantom{XXXXXXXXXXXX}
    \phantom{XXXXXXXXXXXX}
    \textrm{deviation:}
  }
  \begin{CD}
    \Psi_0{=}\Ical @>{\Psi_1}>> \Psi_1      \\
    @VV{\Psi}V                     @VV{\Psi'}V \\
    \Psi              @>>>         \Pst
  \end{CD}
  \qquad\qquad\qquad
  \begin{CD}
     A_0{=}0 @>{{\rm Ad}_{\Psi_1}}>> A_1                    \\
     @VV{{\rm Ad}_\Psi}V             @VV{{\rm Ad}_{\Psi'}}V \\
     A_0{+}A @>>>                    \At
  \end{CD}
\eeq
where vertical arrows turn on a deviation, again via dressing.
Composing the two dressing transformations
(defining $\Pst=\Ps'\Ps_1$ and $\At=A_1{+}A'$) one gets 
\beq
0\= ( Q + \sfrac{1}{\la}X(\ii)\,\h_0 + \At )\,\Pst
 \= \bigl[ ( Q' + \sfrac{1}{\la}X(\ii)\,\h_0 + A' )\,\Ps'\bigr]\,\Ps_1 
\eeq
with $Q'\Ps':= Q\Ps'+ [A_1,\Ps']$. 
Measuring fields from the new reference field~$A_1$ we obtain
\beq
( Q' + \sfrac{1}{\la}X(\ii)\,\h_0 + A' )\,\Ps' \=0 
\eeq
which takes the same form as (\ref{modkey}), except for $Q\mapsto Q'$.

\section{Tachyon vacuum superstring fields}

Of particular interest is the structure of (super)string field theory 
around the (NS) tachyon vacuum. 
Let us suppose the latter can be reached as~$A_1$ within our ansatz.
It happens to be useful to redefine the fluctuations around~$A_1$ by
a world-sheet parametrization, inducing
\beq
A'\ \mapsto\ \Ucal\,A'\ =:\ \Acal
\qquad\textrm{and}\qquad
\Ps'\ \mapsto\ \Ucal\,\Ps'\ =:\ \ps
\eeq
in such a way that $\Qcal:=\Ucal\,Q'\,\Ucal^{-1}$ is the proper
zero-cohomology pure-ghost ``vacuum'' BRST operator
\cite{Gaiotto:2001ji,Ohmori:2002kj}.
Note that $\h_0$ and $\xi(\ii)$ remain unchanged.
Thus, our key equation for vacuum superstring field theory reads
\beq
( \Qcal + \sfrac{1}{\la}\Xcal(\ii)\,\h_0 + \Acal )\,\ps(\la) \= 0
\qquad\textrm{with}\qquad
\Xcal(\ii) \= \{\Qcal,\xi(\ii)\} \quad.
\eeq
Its solutions have the by now familiar form and fulfill the cubic equation
\beq \label{eomvssft}
\Qcal\Acal + \Acal^2\=0 \qquad\textrm{with}\qquad
\Xcal(\ii)\,\h_0\,\Acal\=0 \quad.
\eeq

Assuming~\cite{Rastelli:2000hv} that the vacuum string fields 
describing D-branes factorize, 
\beq
\Acal\=\Acal_g\otimes\Acal_m \qquad\textrm{and}\qquad
\P\=\P_g\otimes\P_m \quad,
\eeq
the field equation~(\ref{eomvssft}) splits into
\beq
\Acal_m^2 \= \Acal_m \qquad\textrm{and}\qquad
\Qcal\Acal_g + \Acal_g^2 \= 0 \qquad\textrm{with}\qquad 
\Xcal(\ii)\,\h_0\,\Acal_g \= 0 \quad.
\eeq
The matter part~$\Acal_m$ is only a ``spectator'' in the linear system.
Our previous analysis then reduces to the solution for the ghost part,
\beq \label{ghostsol}
\Acal_g\=-\sfrac{1{+}\m\mb}{\m}\Xcal(\ii)\,\h_0 P_g
\qquad\textrm{with}\qquad
(\Ical{-}P_g) (\Qcal - \mb\,\Xcal(\ii)\,\h_0) P_g \=0
\eeq
for a ghost projector~$P_g$.
Three remarks are in order.
First, neither $\ps$ nor $\e^{\pm\P}$ factorize.
Second, the form~(\ref{ghostsol}) is {\it not\/} compatible with
the ansatz $\P_m^2=\P_m$ proposed by~\cite{Marino:2001ny}.
Third, nontrivial solutions to~(\ref{ghostsol}) are not obtained via 
$\Qcal P_g=0$ but rather governed by the cohomology 
of the operator~$\ \Qcal-\mb\,\Xcal(\ii)\,\h_0\ $ 
for a given moduli parameter~$\m$.

Let us briefly expand on the last remark. From
\beq
\h_0\,\ps(\la) \= 0 \qquad\Longrightarrow\qquad 
\h_0\Phi = 0 \quad \textrm{and}  \quad \h_0 P = 0
\eeq
we infer that $(\Ical{-}P)\Qcal P=0$ and indeed~$\Acal=0$.
Thus, the simplest situation of $\Qcal$-closed projectors in the
``small Hilbert space'' cannot describe D-branes~\cite{Ohmori:2002ah}.
Such ``supersliver states'' can be expressed as squeezed 
(i.~e.~generalized coherent) one-string states over the first-quantized vacuum
and have been constructed recently~\cite{Arefeva:new}.
For a solution featuring a non-vanishing~$A$, however, one must construct
a ghost projector which satisfies (\ref{ghostsol}) in a less trivial manner,
e.~g., by intertwining its $(b,c)$ and $(\b,\g)$ ghost contents 
in an appropriate fashion.

\section{Outlook}

The ``linearization'' of the superstring field equations provides us with
a new window to nonperturbative superstring physics. The ideas presented here
are just the beginning. Among future developments one may consider
\begin{itemize}
\item the construction of nontrivial classical string fields ($\Acal{\neq}0$)
\item the precise relation with noncommutative solitons (in the Moyal basis)
\item the generalization to the multi-pole ansatz for $\Ps(\la)$
\item the interpretation of solutions (as D-branes?) and their moduli~$\m$
\item the computation of energy densities, tensions, etc.~for a given solution
\item the analysis of fluctuations around constructed classical configurations
\item the extension to the Ramond sector
\end{itemize}
First steps have been made in~\cite{Kling:2002ht}.


\bigskip\noindent
{\bf Acknowledgement} 

\noindent
O.~L.~would like to thank the organizers for a very stimulating conference
in a pretty setting.  
He also acknowledges discussions with I.~Aref'eva and L.~Bonora.


\end{document}